\documentclass[%
 reprint,
superscriptaddress,
showpacs,preprintnumbers,
 amsmath,amssymb,
 aps,
]{revtex4-2}

\usepackage{graphicx}
\usepackage{dcolumn}
\usepackage{bm}
\usepackage{amsmath,mathdots} 
\usepackage{changes}
\usepackage{cancel}
\usepackage{ulem}
\usepackage{comment}

\usepackage[pagebackref=false,colorlinks,linkcolor=blue,filecolor=green,urlcolor=blue ,citecolor=blue]{hyperref}

\usepackage{xcolor}

\begin{document}

\title{Topological phases of commensurate or incommensurate non-Hermitian Su-Schrieffer-Heeger lattices}

\author{Milad Jangjan}
\affiliation{Department of Physics, Faculty of Science, University of Zanjan, Zanjan 45371-38791, Iran}

\author{Linhu Li}
\affiliation{Guangdong Provincial Key Laboratory of Quantum Metrology and Sensing and School of Physics and Astronomy,Sun Yat-Sen University (Zhuhai Campus), Zhuhai 519082, China}

\author{Luis E. F. Foa Torres}
\affiliation{Departamento de F\'{\i}sica, Facultad de Ciencias F\'{\i}sicas y Matem\'aticas, Universidad de Chile, Santiago, 837.0415, Chile}

\author{Mir Vahid Hosseini}
\email[Corresponding author: ]{mv.hosseini@znu.ac.ir}
\affiliation{Department of Physics, Faculty of Science, University of Zanjan, Zanjan 45371-38791, Iran}

\begin{abstract}
We theoretically investigate topological features of a one-dimensional Su-Schrieffer-Heeger lattice with modulating non-Hermitian on-site potentials containing four sublattices per unit cell. The lattice can be either commensurate or incommensurate. In the former case, the entire lattice can be mapped by super cells completely. While in the latter case, there are two extra lattice points, thereby making the last cell incomplete. We find that an anti-$\mathcal{PT}$ transition occurs at exceptional points of edge states at certain parameters, which does not coincide with the conventional topological phase transition characterized by the Berry phase, provided the imaginary on-site potential is large enough. Interestingly, when the potential exceeds a critical value, edge states appear even in the regime with a trivial Berry phase. To characterize these novel edge states we present topological invariants associated with the system's parity.
Finally, we analyze the dynamics for initial states with different spatial distributions, which exhibit distinct dynamics for the commensurate and incommensurate cases, depending on the imaginary part of edge state energy.
\end{abstract}

\maketitle

\section {Introduction} \label{s1}

The systematic examination of non-Hermitian (NH) Hamiltonians with $\mathcal{PT}$ symmetry was started by Bender and Boettcher in their pioneering work \cite{Bender}. They established that a broad class of $\mathcal{PT}$-symmetric Hamiltonians possesses real spectra \cite{RealSpectra1,RealSpectra2,RealSpectra3,RealSpectra4}, a remarkable departure from the complex eigenvalues typically associated with NH systems. This discovery has paved the way for a deeper investigation of the intriguing connection between symmetry and topology within the realm of NH physics \cite{non-HermPhys1,non-HermPhys2,non-HermPhys3,non-HermPhys4}.

Recent research has brought to the forefront the significance of NH analogs of the Su-Schrieffer-Heeger (SSH) model \cite{SSH} when it comes to elucidating one-dimensional (1D) topological behavior within systems characterized by gain and/or loss \cite{PT6,PT7,PT8,PT9,PT10,PT11,PT12}. This intriguing avenue of study has garnered attention due to its potential to uncover unconventional phenomena in condensed matter physics. In particular, a multitude of theoretical investigations have put forth innovative approaches to engineer versions of the SSH model that exhibit $\mathcal{PT}$-symmetry \cite{PT1,PT2,PT3,PT4,PT5,PT6,PT7,PT8,PT9,PT10,PT11,PT12,PT13} or anti $\mathcal{PT}$-symmetry \cite{Anti_PT1,Anti_PT2}. These efforts extend our understanding of topological behavior beyond NH systems, delving into the intricate interplay between gain and loss. 

In the study of topological phases in condensed matter physics, especially in Hermitian \cite{Topo_Hermit1,Topo_Hermit2,Topo_Hermit3,Topo_Hermit4,Topo_Hermit5,Topo_Hermit6,Topo_Hermit7,Topo_Hermit8,Topo_Hermit9,Topo_Hermit10,Topo_Hermit11,Topo_Hermit12} and NH systems \cite{Topo_Non_Hermit1, Topo_Non_Hermit2,Topo_Non_Hermit3,Topo_Non_Hermit4,Topo_Non_Hermit6,Topo_Non_Hermit7,Topo_Non_Hermit8,Topo_Non_Hermit9,Topo_Non_Hermit10,Topo_Non_Hermit11,Topo_Non_Hermit12,Topo_Non_Hermit13,Topo_Non_Hermit14,Topo_Non_Hermit15,Topo_Non_Hermit17,Topo_Non_Hermit18,Topo_Non_Hermit19,Topo_Non_Hermit20,Topo_Non_Hermit21,Topo_Non_Hermit22,Topo_Non_Hermit23, Topo_Non_Hermit24,Topo_Non_Hermit25,SkinEffect1,SkinEffect2,SkinEffect3}, the role of symmetry is paramount, and inversion symmetry is no exception. To delve deeper into the intricate interplay between inversion symmetry and topological edge states, we turn our attention to a 1D lattice system where inversion symmetry and bulk-translational symmetry are incompatible. This intriguing scenario leads to profound consequences, shedding light on the behavior of edge states in both topological and non-topological phases. Traditionally, the coexistence of bulk-translational symmetry and inversion symmetry has been a cornerstone of the bulk-boundary correspondence principle, which stipulates that the topological properties of a system are reflected in the presence or absence of edge states. However, in the absence of bulk-translational symmetry, this correspondence appears to be challenged. For example, the celebrated SSH model with an odd number of lattice sites exhibits one spatially asymmetric edge state in both its topological nontrivial and trivial phases. This sophisticated behavior highlights the intricate interplay between different symmetries in determining the topological properties of a system. 

In this intriguing landscape, inversion symmetry emerges as a savior of symmetry-based topological protection \cite{Inv_sym1, Inv_sym2,Inv_sym3}. Even in NH systems, where the preservation of Hermiticity is not guaranteed, inversion symmetry plays a crucial role \cite{Inv_Sym_No-Her1,Inv_Sym_No-Her2,Inv_Sym_No-Her3}. Specifically, it ensures that the edge states residing at opposite ends of the 1D chain possess identical imaginary energies. This remarkable feature means that these edge states share the same decaying rate during time evolution, preserving their spatial symmetry. Consequently, in parameter regimes where edge states exhibit the largest imaginary energies, inversion symmetry acts as a guardian, protecting the spatially symmetric edge localization over extended periods of time. This phenomenon in NH systems underscores the profound influence of inversion symmetry on the topological and non-topological edge states, enriching our understanding of the interplay between symmetry and topology in condensed matter systems.

Here, we explore the characteristics of a NH chain model with four sublattices. While various models consisting of four sublattices have been extensively investigated \cite{Anti_PT1,Anti_PT2}, it is noteworthy that in this model, anti-$\mathcal{PT}$ symmetry protected edge states only manifest themselves within the open boundary conditions (OBCs), with no trace on periodic boundary conditions (PBCs). In addition, our model is distinctive in its flexibility, with parameters that accommodate both real and imaginary values, and the possibility to explore the effect of the number of unit cells, in the "commensurate case" with a full number of unit cells, inversion symmetry is disrupted, while the "incommensurate case" with an integer number of unit cells plus two extra lattice sites - restores inversion symmetry at the cost of translational symmetry. We utilize the twist boundary condition to delineate phase separation, a strategy contrasting with previous research ~\cite{Anti_PT1} that emphasized anti-$\mathcal{PT}$ symmetry's role in distorting the imaginary components of edge state energies. Instead, we propose that inversion asymmetry primarily drives this distortion.

A key focus of our research is the role of exceptional points (EPs)~\cite{Ep1,Ep2,Ep3} in NH systems. These points, lacking parallels in Hermitian systems, introduce distinct topological features that contrast with Hermitian degeneracy points. Through our analysis, we discern both exceptional points and topological phase transition points at specific potential intensities, a task unachievable by the Berry phase. The concept of the berry phase is intricately tied to the closure and subsequent reopening of bulk states. However, it is important to note that within certain regions, there exist degenerate edge states that are not necessarily linked to the closure and reopening of bulk states. These edge states emerge from EPs within open boundary conditions and are protected by anti-$\mathcal{PT}$ symmetry and inversion symmetry. These degenerate edge states represent a distinct phenomenon, highlighting the new kind of topological phases in materials. This insight leads us to develop a new interpretation grounded in the concept of inversion symmetry, to differentiate between these transition types. Moreover, we delve into the dynamic facets of these phenomena, enhancing our comprehension of topological behaviors in NH systems.


The paper is organized as follows. In Sec. \ref{s2}, we present the Hamiltonian of the system and reveal its symmetries. Section \ref{s3} presents the obtained numerical results for band structures and the Berry phase. Also, the quantification of non-Hermiticity and developing new topological invariants are discussed in Sec. \ref{s4}. In Sec. \ref{s5}, the dynamics of both bulk and edge states are analyzed in different lattice termination. Finally, Sec. \ref{s6} is devoted to concluding remarks.

\section {Model and Theory}\label{s2}

\begin{figure}[t!]
    \centering
    \includegraphics[width=1\linewidth]{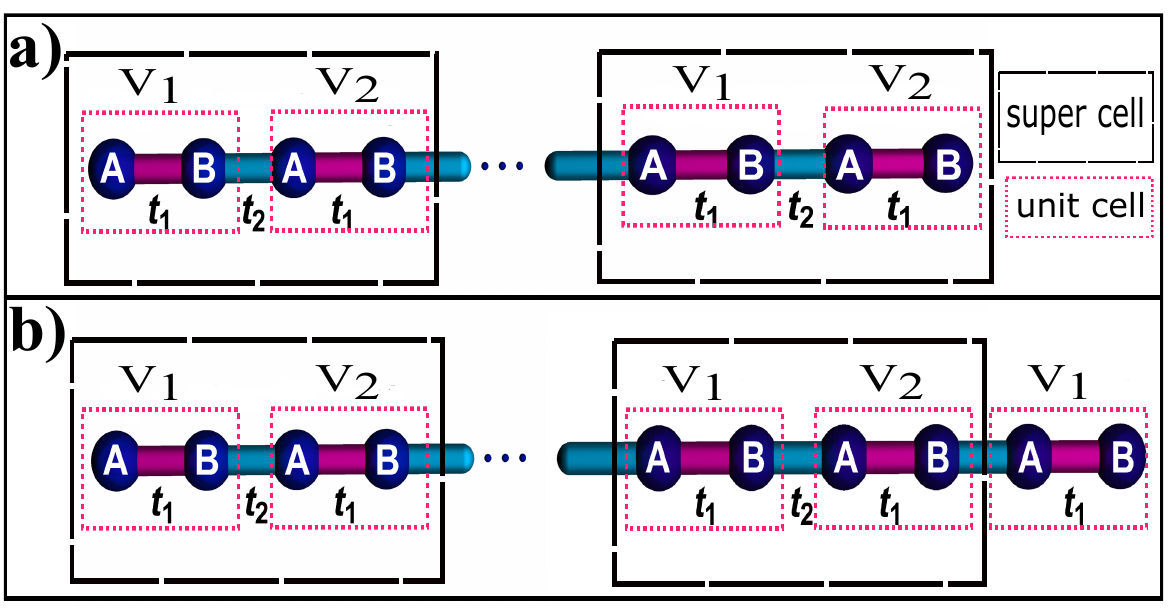}
    \caption{(Color online) Schematic representations of (a) commensurate and (b) incommensurate SSH lattice with alternating hopping energies ($t_1$ and $t_2$) and two (four) sublattices per unit (super) cell. The odd unit cells experience the on-site potential $V_1$ and the even unit cells have the on-site potential $V_2$.}
    \label{fig1}
\end{figure}

We consider a 1D NH SSH model with commensurate and incommensurate lengths as shown in  Figs. \ref{fig1}(a) and \ref{fig1}(b), respectively. The total tight-binding Hamiltonian of the systems is comprised of the SSH lattice ($H_0$) and on-site potential ($U$), given by 
\begin{eqnarray}\label{s2E1}
H=H_0+U,
\end{eqnarray}
where 
\begin{eqnarray}\label{s2E2}
H_0&=&\sum_{i=1}^{n}t_1 A_{i} ^\dagger B_{i}+\sum_{i=1}^{n-1} t_2A_{i+1}^{\dagger} B_{i}+h.c,\nonumber\\ 
U&=&\sum_{i=1}^{n}\sum_{l=1}^{n} (V_1 \delta_{i,2l-1}+V_2 \delta_{i,2l})(A_{i} ^\dagger A_{i}+B_{i} ^\dagger B_{i}),\nonumber
\end{eqnarray}
where $n$ is the number of unit cells and $\delta_{i,j}$ is the Kronecker delta function. Here, $X^{\dagger}_{i}$ ($X_{i}$), $X\equiv(A,B)$, is the creation (annihilation) operators on the sublattices $A$ and $B$ at the i{\it th} unit cell. $t_1 =t[1+\delta_0 \cos(\theta)]$ and $t_2 =t[1-\delta_0 \cos(\theta)]$ are the modulated hopping amplitudes with a phase factor $\theta$. Throughout this paper, the modulation intensity describing the strength of dimerization is chosen to be $\delta_0=0.8$, and $t=1$ is set to be the energy unit. In addition, $V_1 (V_2)$ stands for the on-site potential in the odd (even) unit cells, defined as 
\begin{eqnarray}
V_1=\gamma_1 e^{i\alpha}, ~~V_2=\gamma_2 e^{i\alpha}, \label{s2E3}
\end{eqnarray}
where $\gamma_1 (\gamma_2)$ is the strength of the potential, and $\alpha$ is the phase of the potentials resulting in $V_1 $ and $V_2$ can be either real or imaginary depending on the values of $\alpha$. 

The case of two-sublattice unit cell in 1D systems similar to the original SSH model \cite{SSH} is well-studied \cite{SSH2,GeneralizedSSH,SSHSpinOrbit,SpinZem}. Remarkably, unlike the previous studies, in the presence of $U$, since, the period of on-site potential modulation is twice the dimerization period of the lattice, one must enlarge the unit cells \cite{enlargeUnit} such that the established super cells include four distinct sublattices. Consequently, if $n$ is an even number, the lattice can be mapped by a four-sublattice supercell ($4N$), leading to a commensurate lattice. Conversely, for odd values of $n$, one deals with an incommensurate case, where, in addition to the sublattices placed in the supercells, the chain contains two additional lattice sites ($4N+2$). So, this leads to breaking the transitional symmetry in the presence of $U$. Here, $N$ is the number of supercells.

For the commensurate case, under PBCs, one can take Fourier transformation and use the momentum space Hamiltonian to obtain the topological phase transition points. Under PBCs, the Hamiltonian with even number of unit cells can be expressed in the momentum space as
\begin{eqnarray}\label{s2E4}
H=\sum_k \psi_k^\dagger h(k) \psi_k,
\end{eqnarray}
where the four-component basis vector is $\psi_k=(A_{1k},B_{1k},A_{2k},B_{2k})^\dagger$, and the momentum space Hamiltonian $ h(k)$ reads
\begin{eqnarray}\label{s2E5}
 h(k)=
 \left ( \begin{array}{c c c c}
     V_1 & t_1 & 0 & t_2 e^{ik} \\
     t_1 & V_1 & t_2 & 0 \\
      0 & t_2 & V_2 & t_1 \\
     t_2 e^{-ik} & 0 & t_1 & V_2
 \end{array}\right)
\end{eqnarray}
\begin{eqnarray}
&=&\frac{(V_1-V_2)}{2}\sigma_z\otimes\sigma_0+\frac{(V_1+
V_2)}{2}\sigma_0\otimes\sigma_0+t_1\sigma_0\otimes\sigma_x\nonumber\\
&&+\frac{t_2(1+\cos k)}{2}\sigma_x\otimes\sigma_x+\frac{t_2(1-\cos k)}{2}\sigma_y\otimes\sigma_y\nonumber\\
&&+\frac{t_2\sin k}{2}(\sigma_y\otimes\sigma_x+\sigma_x\otimes\sigma_y),
\end{eqnarray}
with $k$ is the momentum, and $\sigma_0$ and $\sigma_{x,y,z}$ are the two-by-two identity matrix and Pauli matrices, respectively. 
In particular, if we set $\alpha=\pi/2$ and $\gamma_1=-\gamma_2=\gamma$, so that $V_1=-V_2=V=i\gamma$ are purely imaginary.
Diagonalzing the above Hamiltonian yields the following four energy bands,
\begin{eqnarray}\label{s2E6}
E_{\pm,\pm}=\pm\frac{\sqrt{X \pm \sqrt{Y}}}{2},
\end{eqnarray}
where
\begin{eqnarray}\label{s2E7}
X&=&4(t_1^2+t_2^2)-4\gamma^2, \nonumber \\
Y&=&16t_1^2[2t_2^2(1+\cos(k))-4\gamma^2].
\end{eqnarray}
In the presence of the imaginary potential, the NH system described by Eq. (\ref{s2E5}) has a transposition time-reversal symmetry ($TRS^{\dagger}$) and a conjugate particle-hole symmetry ($PHS^{\star}$) \cite{PhysRevX.9.041015,Classification}, represented by
$\mathcal{T} h^{T}(k) \mathcal{T}=h(-k)$ and $\mathcal{C} h^{*}(k) \mathcal{C}=-h(-k)$, respectively, with operators $\mathcal{T}=\sigma_x \otimes \sigma_x$, and $\mathcal{C}=\sigma_0 \otimes \sigma_z$. Consequently, $h(k)$ has chiral symmetry (pseudo-anti-non-hermiticity) defined by $\Gamma h^{\dagger}(k) \Gamma=-h(k)$ with $\Gamma=\mathcal{T} \mathcal{C}$, and belongs to the $BDI^{\dagger}$ class in the 38-fold topological classifications of NH systems \cite{PhysRevX.9.041015}. In addition, Eq. (\ref{s2E5}) satisfies a $\mathcal{PT}$ symmetry as 
\begin{eqnarray}
(\mathcal{PT})h^*(k)(\mathcal{PT})^{-1}=h(k),
\end{eqnarray}
with $\mathcal{PT}=\sigma_x \otimes \sigma_x$, which can be broken for all eigenstates when $Y<0$ $\forall k$, i.e.
\begin{eqnarray}\label{e1212}
2t_2^2(1+\cos (k))-4\gamma^2<0 \Rightarrow \gamma^2>t^2_2.
\end{eqnarray}
However, under OBCs, the $\mathcal{PT}$ symmetry is satisfied only for the commensurate case which the operator for such symmetry can be represented as
\begin{eqnarray}\label{e3} \mathcal{PT}&=&\sigma_{X_{4N}}=\begin{pmatrix}
 & &  & &1\\
  & O& &1 &\\
  & &\iddots & & \\
   & 1&  &O & \\
   1 & & & & \\
 \end{pmatrix}_{4N}.
\end{eqnarray}

In contrast to the PBCs, where anti-$\mathcal{PT}$ symmetry is absent, it is noteworthy that Hamiltonian (\ref{s2E1}) exhibits the anti-$\mathcal{PT}$ symmetry for both the commensurate and incommensurate cases as $(\mathcal{APT})H^*(\mathcal{APT})^{-1}=-H$ with an explicit forms given by $\mathcal{APT}=\sigma_{0_{n}} \otimes \sigma_z$ where $\sigma_{0_{n}}$ is $n$-by-$n$ identity matrix. The presence and absence of anti-$\mathcal{PT}$ symmetry, respectively, under OBCs and PBCs imply that zero-real energy edge states will be affected by this symmetry.
\begin{figure}[t!]
    \centering     
    \includegraphics[width=1\linewidth]{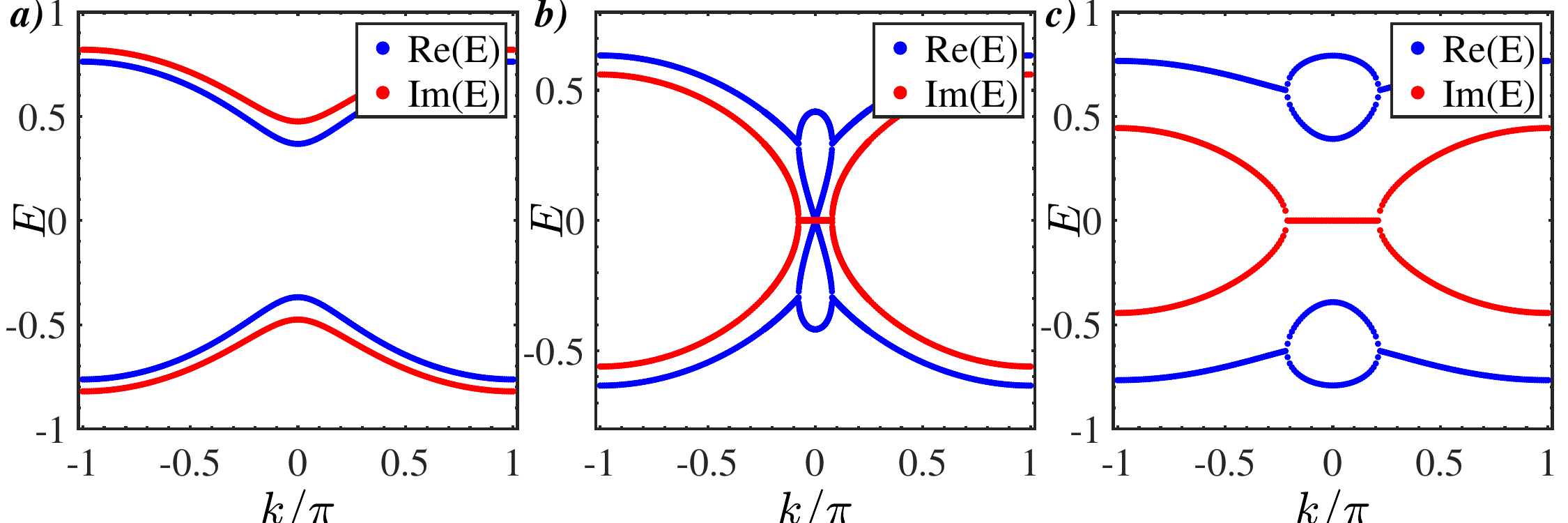}
    \caption{(Color online) The PBC band structure for the model with imaginary on-site potentials $\gamma=1.7$., with (a) $\theta/\pi=0.8$($|\gamma|>t_2$), 
    (b) $\theta/\pi\approx0.86$($|\gamma|<t_2$, $t_1=\sqrt{t_2^2-\gamma^2}$), 
    and (c) $\theta/\pi=1$($|\gamma|<t_2$).}
    \label{fig5}
\end{figure}

In Hamiltonian (\ref{s2E5}), a topological phase transition can be determined by the gap closing condition $X^2=-Y$, which is satisfied only at $k=0$ with
\begin{eqnarray}\label{s2E8}
t_1=\pm \sqrt{t_2^2-\gamma^2},
\end{eqnarray}
provided that all parameters ($t_1$, $t_2$, and $\gamma$) take nonzero real values. Also, the other gap closing condition $Y=0$ does not signify a topological phase transition associated with zero-real-energy edge states in our model. Instead, it corresponds to the subgap closing between the two subbands with positive (or negative) real energies in Eq. (\ref{s2E6}), which is not our focus since a subgap is not protected by any symmetry, and subbands may overlap each other and become inseparable even away from $Y=0$. Note that this model does not suffer from the NH skin effect \cite{SkinEffect1,SkinEffect2,SkinEffect3}, as it satisfies the spinless time reversal symmetry in its Hermitian limit \cite{SkinEffect4}.
Therefore, the gap closing condition and its associated topological phase transition shall be the same for both PBCs and OBCs, and also the twisted boundary conditions (TBCs) discussed later.

To validate Eqs.  (\ref{e1212}) and (\ref{s2E8}), and highlight the broken and unbroken phases of $\mathcal{PT}$ symmetry in the PBC system, the real and imaginary parts of the bands versus momentum $k$ are plotted in Fig. \ref{fig5} for different parameters. 
In Fig. \ref{fig5}(a), we display an example where Eq. (\ref{e1212}) is satisfied, and it is evident that all eigenenergies take complex values, indicating a $\mathcal{PT}$-broken phase. By adjusting $\theta$ to the value where Eq. (\ref{s2E8}) holds, 
a topological phase transition occurs with band gap closes for both the real and imaginary parts of the spectrum, as displayed in Fig. \ref{fig5}(b). On the other hand, as Eq. (\ref{e1212}) is violated for these parameters, eigenenergies are seen to take real values (with ${\rm Im}[E]=0$) for a certain region of $k$ with unbroken $\mathcal{PT}$ symmetry. In Fig. \ref{fig5}(c), the energy gap reopens with $\theta$ increases, and the energy spectrum possesses $\mathcal{PT}$-broken for large $|k|$ and $\mathcal{PT}$-unbroken phase around small values of $|k|$.

\section{Berry phase}\label{s3}

In the incommensurate case, the TBCs \cite{TwistBC1,TwistBC2,TwistBC3,Kohn_Condensed1,Kohn_Condensed2} is more relevant since $k$ is not a good quantum number to define topological invariants. 
In our model, the coupling $\tau$ between the first and last sites specifies the boundary conditions as
\begin{eqnarray}\label{s2E4}
\tau= \begin{cases}
0 \quad  \quad {\rm OBC} \\ \nonumber
t_2 \quad  \quad {\rm PBC} \\ \nonumber
t_2 e^{i\Theta} \quad {\rm TBC}
\end{cases},
\end{eqnarray}
where the phase $\Theta$, known as torsion, is a real number.
For NH Hamiltonians $H^\dagger\neq H$,
distinct left ($|\Psi_m\rangle$) and right ($|\Phi_{m}\rangle$) eigenstates emerge, satisfying the Schr\"odinger equation as $H^\dagger |\Psi_{m}\rangle = E_m^\star |\Psi_{m}\rangle$ and $H|\Phi_{m}\rangle = E_m|\Phi_{m}\rangle$~\cite{biorthogonal}. Here, $E_{m}$ is the corresponding eigenenergy with $m$ being the index of eigenenergy.  In NH systems, the process of normalizing eigenvectors involves a bilinear product that combines both the left and right eigenvectors, i.e.,
\begin{equation}
|\Tilde{\Psi}_{m}\rangle =\frac{|\Psi_{m}\rangle}{\sqrt{\langle \Phi_m|\Psi_{m}\rangle}}, \nonumber \\
|\Tilde{\Phi}_{m}\rangle =\frac{|\Phi_{m}\rangle}{\sqrt{\langle \Phi_m|\Psi_{m}\rangle}}.\nonumber \\
\end{equation}
Now, the Berry phase under TBCs~\cite{BerryPhase,TBCberry} can be defined as~\cite{NH_berry_phase1,NH_berry_phase2} 
\begin{equation}
\beta = \sum_{m=1}^{N_f} i\int_{0}^{2\pi} \langle \Tilde{\Psi}_{m}(\Theta) | \frac{\partial}{\partial \Theta} | \Tilde{\Phi}_{m}(\Theta) \rangle d\Theta,
\end{equation}
where $N_f$ is the number of bands below the Fermi energy. In what follows, we will discuss topological properties and the band structures of the system for both real ($\alpha=0,\pi$) and imaginary ($\alpha=\pi/2$) on-site potentials $V_{1,2}$. 

In addition, to distinguish the localized states from the extended ones, we evaluate the inverse participating ratio given by $I_E=\sum_{m}|\Tilde{\Psi}_{m}|^4$ \cite{IPR} for each normalized eigenstate $\Psi_{m}$ with eigenenergy $E_{m}$. Typically, localized states and extended states have $I_E \rightarrow 1$ and $I_E \rightarrow 1/N_{\rm total}\approx 0$, respectively, with $N_{\rm total}=2n$ being the total number of lattice sites.

\begin{figure}[t!]
    \centering    \includegraphics[width=1\linewidth]{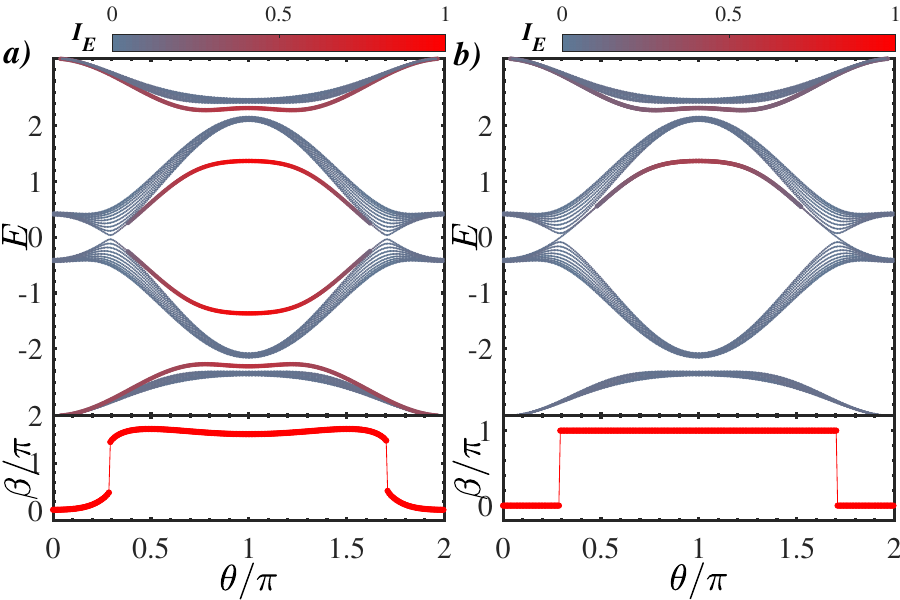}
    \caption{(Color online) Band structure and TBC Berry phase as a function of $\theta/\pi$ for (a) commensurate and (b) incommensurate lattices. Other parameters are $\gamma_1=-\gamma_2=1.4$, and $\alpha=0$. The Fermi energy is set at $E=0$, so that $N_f=2n/2=20$, and $N_f=2n/2=19$ for the TBC Berry phase in panels (a) and (b), respectively.}
    \label{fig2}
\end{figure}

\subsection{Real on-site potentials}

Given the interest in studying the band topology in NH systems, 
for the sake of comparison, we will first present the results in the absence of non-Hermiticity, i.e. with $\alpha=0,\pi$ in Eq. \eqref{s2E3} and thus real on-site potentials. 
With $\gamma_1=-\gamma_2=\gamma$, 
the inversion symmetry of Hamiltonian (\ref{s2E5}) is broken in the commensurate case, but it is preserved in the incommensurate case. The band structure and its corresponding topological invariant (the TBC Berry phase) are shown in Fig. \ref{fig2} for the commensurate [Fig. \ref{fig2}(a)] and incommensurate [Fig. \ref{fig2}(b)] lattices. As can be seen from Fig. \ref{fig2}(a), due to the breaking of the inversion symmetry, two branches of non-degenerate edge states with non-zero energies emerge within the central band gap.
While the band structure manifests the chiral symmetry of $\Gamma$ (reflected by the spectrum symmetric about $E=0$),
the topological invariant does not have a quantized value in the presence of these edge states. On the contrary, for the incommensurate lattice, the inversion symmetry of the system is restored, but the chiral symmetry is broken by the incommensurability. Consequently, edge states emerge with two-fold degeneracy at non-zero energies, corresponding to the quantization of the TBC Berry phase at $\pi$, as shown in Fig. \ref{fig2}(b).

\subsection{Imaginary on-site potentials}

\begin{figure}[t!]
    \centering     \includegraphics[width=1\linewidth]{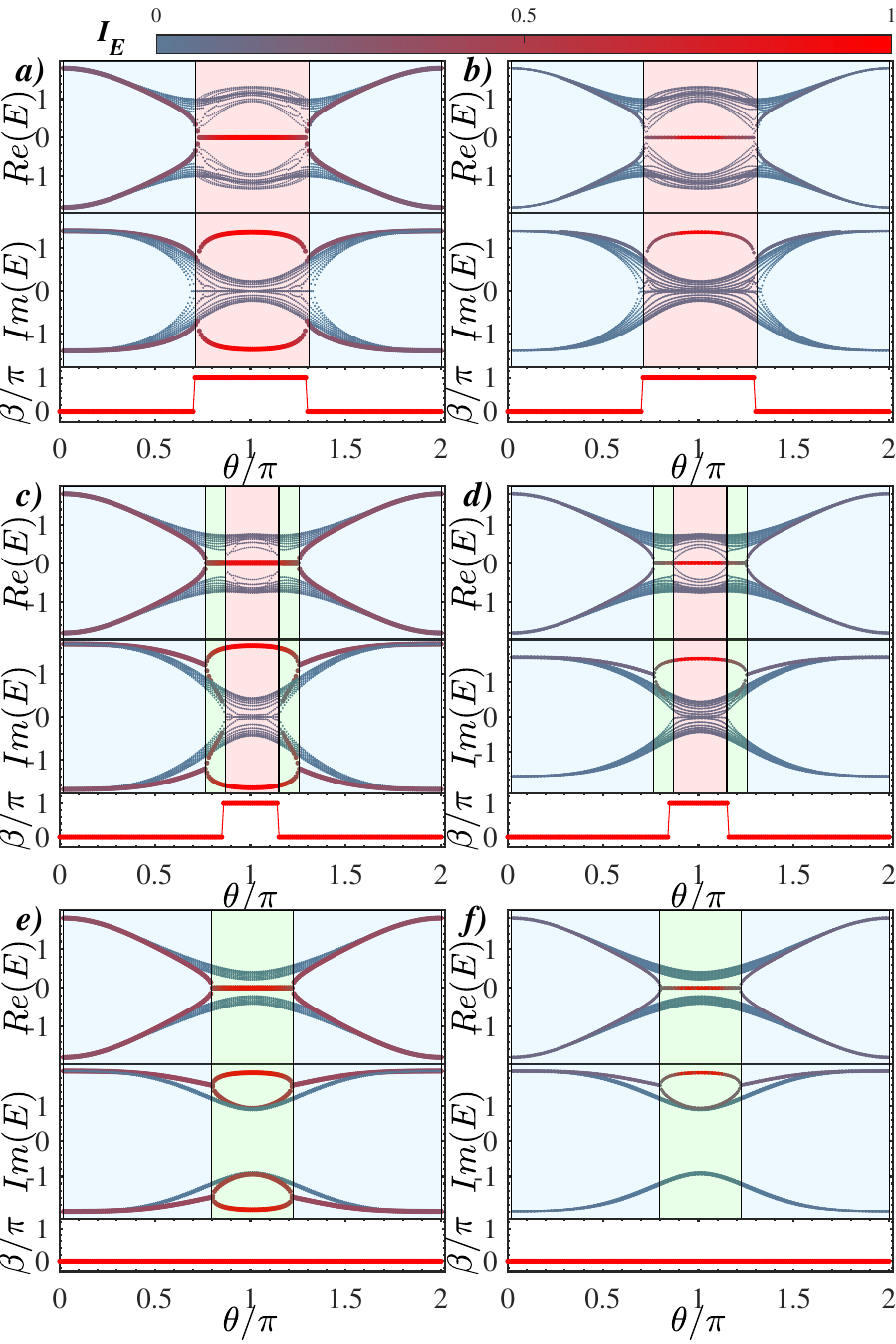}
    \caption{(Color online) Band structure and relevant topological invariant for commensurate (left column) and incommensurate (right column) lattice when (a,b) $\gamma=1.4$, (c,d) $\gamma=1.7$, and (e,f) $\gamma=2$.}
    \label{fig4}
\end{figure}

In the presence of imaginary on-site potentials, we will find that different types of edge states emerge, associated with both the topological phase transition and the symmetries discussed in the previous sections. In Fig. \ref{fig4}, the band structure and the TBC Berry phase with different values of $\gamma$ are shown for the commensurate (left panels) and incommensurate (right panels) cases. In Figs. \ref{fig4}(a) and \ref{fig4}(b), one can see that zero-real-energy edge states emerge in the red region centered at $\theta=\pi$, and merge into the bulk bands after the topological phase transition characterized by the TBC Berry phase $\beta$. The eigenenergies of these edge states form a complex conjugated pair for the commensurate case due to $\mathcal{PT}$ symmetry, and have a two-fold degeneracy in the imaginary part for the incommensurate case owing to inversion symmetry.
The latter case will be shown below. On the other hand, the zero-real-energy of these edge states reflects both the topological protection and the unbroken anti-$\mathcal{PT}$ symmetry.

By increasing the amplitude of the on-site potentials, we observe a separation between the topological phase transition and the vanishing of edge states, as shown in Figs. \ref{fig4}(c) and  \ref{fig4}(d). In particular, the edge states do not merge into the bulk when $\theta$ changes from the red region with $\beta=\pi$ to the green region with $\beta=0$. Instead, two additional zero-real-energy edge states emerge after the topological phase transition. As $\theta$ further moves away from $\pi$, these four edge states divide into two pairs, each possessing the same imaginary energy but opposite real energies, representing an anti-$\mathcal{PT}$-broken phase.
The transition between the anti-$\mathcal{PT}$-broken and -unbroken phases are thus marked by an EP of the edge states.

Finally, with further enlarged amplitude of the on-site potentials, the system becomes always topologically trivial regarding the TBC Berry phase ($\beta=0$), and supports only the anti-$\mathcal{PT}$-broken (blue region) and -unbroken (green region) phases for different $\theta$, hosting anti-$\mathcal{PT}$ edge states with opposite and zero real-energies, respectively (See Figs. \ref{fig4}(e) and \ref{fig4}(f)).

\begin{figure}[t!]
    \centering     
   \includegraphics[width=.8\linewidth]{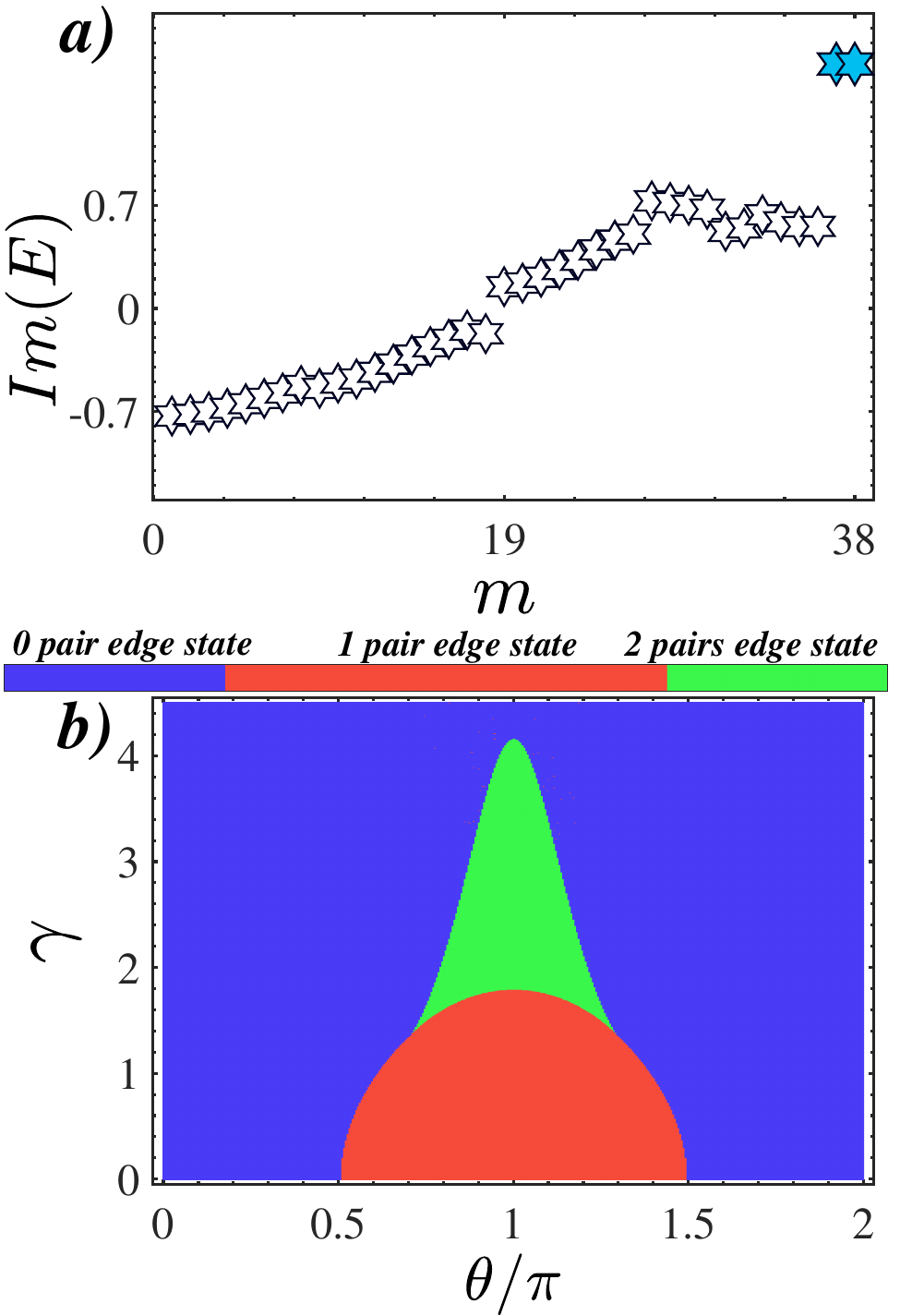}
    \caption{(Color online) (a) Imaginary part of the band structure as a function of eigenenergy index for the incommensurate case with $\gamma=1.7$, $\theta/\pi=1$, and $\zeta=0.5$. The degenerate energy states are represented by the blue stars. (b) Phase diagram of the system in the ($\theta/\pi,\gamma$) plane. The blue color indicates a trivial phase with zero pair of edge states. The red (green) color shows a region where there is one (two) pair(s) of zero-real-energy edge states. Here $\alpha=\pi/2$.}
    \label{fig3}
\end{figure}

In order to demonstrate the role of inversion symmetry on the degeneracy of the imaginary part of the edge states, we modify the on-site potential in the odd unit cells, as $V_1 \rightarrow V_1+\zeta$ where $\zeta$ has an arbitrary real value. This disrupts the anti-$\mathcal{PT}$ symmetry and preserves the inversion symmetry. As illustrated in Fig. \ref{fig3}(a), the effect of potential modification is shown by plotting the imaginary part of band structure versus eigenenergy index. One can see the degeneracy of the imaginary part (blue stars) is preserved in the absence of anti-$\mathcal{PT}$ symmetry. This emphasizes the role of inversion symmetry in protecting the degeneracy of the imaginary component.

By counting the number of zero-real-energy edge states, we have constructed a phase diagram as functions of $\theta$ and $\gamma$ for the system, as depicted in Fig. \ref{fig3}(b). This phase diagram distinctly presents three significant regions within the system's parameter space. Topological phase transitions characterized by the TBC Berry phase are indicated by the boundary between the red region (with one pair of zero-real-energy edge states) and the rest. For $|\gamma|$ exceeding a critical value ($\gamma_c\sim1.4$), an anti-$\mathcal{PT}$ transition emerges, marked by the boundary between green (with two pairs of zero-real-energy edge states) and blue (without zero-real-energy edge states) regions.

\begin{figure*}[t!]
    \centering     \includegraphics[width=1\linewidth]{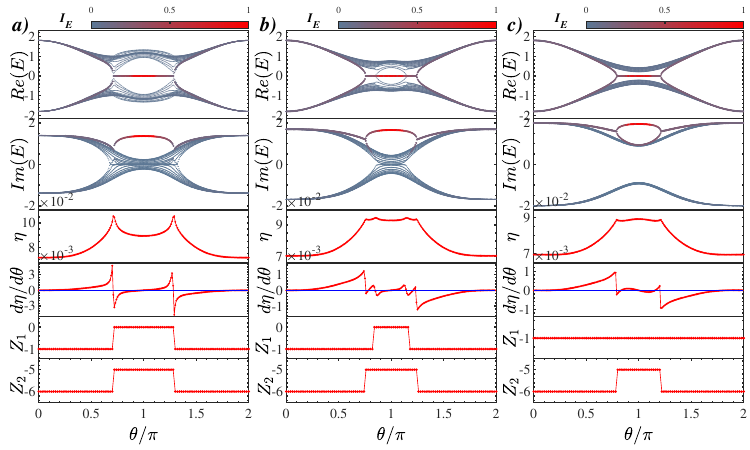}
    \caption{(Color online) Band structure and relevant topological invariant and also $\eta$ and $d\eta/dt$ as the function of $\theta/\pi$ for (a) $\gamma=1.4$ (b) $\gamma=1.7$ and (c) $\gamma=2$.}
    \label{fig6}
\end{figure*}

\section {non-Hermitian quantification and Z invariant in real space} \label{s4}

As already shown above, the anti-$\mathcal{PT}$ edge states are a consequence of the non-Hemiticity of the system, and its phase boundary is not accompanied by a topological phase transition of $\beta$. As such, the TBC Berry phase characterizing the gap closing/reopening cannot determine the topology of the corresponding phases appropriately. Therefore, in the following discussion, we will first quantify the non-Hermiticity of the system, and then introduce different symmetry-based topological invariants to characterize and discriminate the edge states of the system.

The utilization of bi-orthogonal bases allows us to investigate the intricacies of NH quantum systems by establishing a framework within the Hilbert space. However, the challenge lies in devising a concise and efficient method to characterize the departure from the Hermitian system, due to the non-orthogonality among right eigenvectors (or left eigenvectors). To address the issue of the lack of Hermiticity of NH systems, our focus is limited to an in-depth analysis of properties that are exhibited solely by the right basis. This approach is taken instead of an examination of the entire bi-orthogonal basis, given the analogous characteristics shared by the right eigenvectors. A quantification measure designed to specify the degree of non-Hermiticity can be defined as \cite{Eta1,Eta2} 
\begin{eqnarray}
    \eta=\frac{\sum_{m^{\prime}<m} |\langle \Tilde{\Phi}_{m^{\prime}}|\Tilde{\Phi}_m\rangle|^2}{\sum_{m^{\prime}<m}|\langle \Tilde{\Phi}_{m^{\prime}}|\Tilde{\Phi}_{m^{\prime}}\rangle||\langle \Tilde{\Phi}_m|\Tilde{\Phi}_m\rangle|},
\end{eqnarray}
where $\eta=0$, all $|\Tilde{\Phi}_{m^{\prime}}\rangle$ are mutually orthogonal to each other, corresponding to unitary systems (Hermitian or anti-Hermitian),
and $\eta=1$, all eigenvectors coalesce into a single one, representing the case of non-unitarity.

In Figs. \ref{fig6}(a), \ref{fig6}(b), and \ref{fig6}(c), we illustrate the behavior of $\eta$, together with the corresponding band structures, for our model with $\gamma=1.4$, $1.7$, and $2$, respectively. It is seen that $\eta$ reaches a local maximum at both the topological phase transition (indicated by the bulk gap closing points) and the anti-$\mathcal{PT}$ transition (indicated by the edge EPs). In addition, the first-order derivative $\partial \eta$ becomes discontinuous at the anti-$\mathcal{PT}$ transitions, reflecting the emergence of EPs [which coincide with the topological phase transition in Fig. \ref{fig6}(a)]. Note that in contrast to Ref. \cite{Eta1}, the edge EPs in our model do not correspond to a transition in the orthogonality and localizing direction of edge states, and hence do not induce discontinuity to $\eta$ itself.

\begin{figure}[t!]
    \centering     \includegraphics[width=1\linewidth]{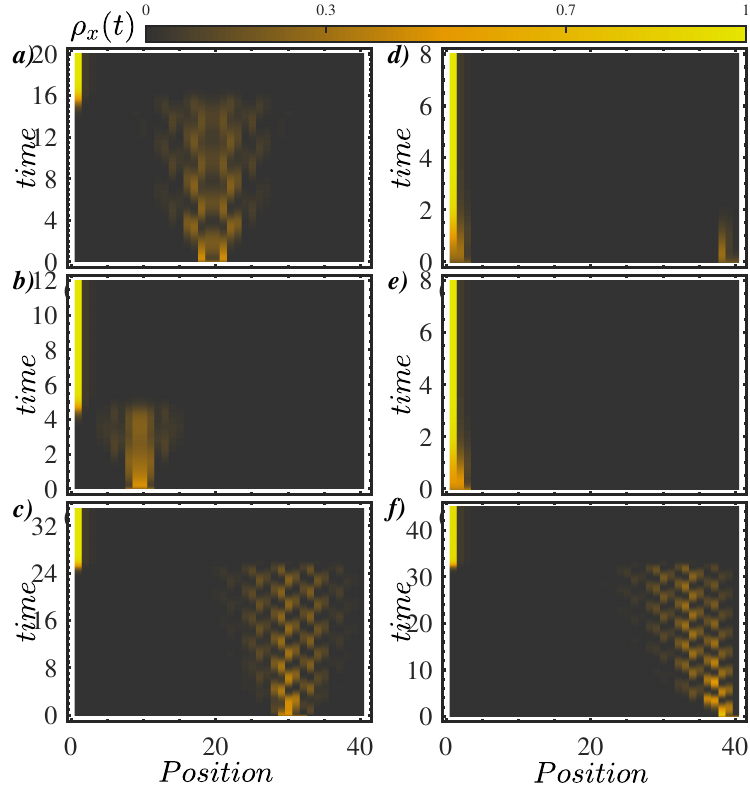}
    \caption{(Color online) Time evolution of different bulk (left) and edge (right) states for the commensurate case. The initial bulk state is prepared at (a) the center, (b) the middle of the left half, and (c) the middle of the right half of the chain. The initial edge state is prepared at (d) both edges, (e) left edge, and (f) right edge of the system. Parameters are $\gamma=1.7$ and $\theta=\pi$.}
    \label{fig8}
\end{figure}

\begin{figure}[t!]
    \centering     \includegraphics[width=1\linewidth]{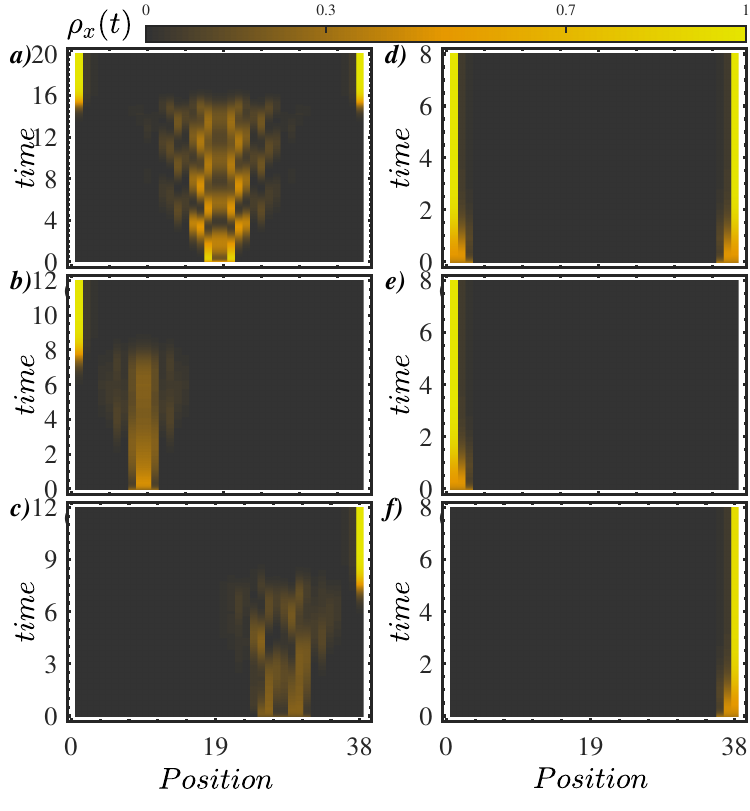}
    \caption{(Color online)  Time evolution of different bulk (left) and edge (right) states for the incommensurate case. The initial bulk state is prepared at (a) the center, (b) the middle of the left half, and (c) the middle of the right half of the chain. The initial edge state is prepared at (d) both edges, (e) left edge, and (f) right edge of the system. Parameters are $\gamma=1.7$ and $\theta=\pi$.}
    \label{fig9}
\end{figure}

Taking into account the inversion symmetry for our model in the incommensurate case,
we further propose to characterize the edge states by counting the parities of all OBC eigenstates below the Fermi energy (i.e. with ${\rm Re}[E]<0$).
Explicitly, the presence of inversion symmetry ensures that the commutation of the Hamiltonian with the inversion symmetry operator yields $[H,\Pi]=0$ under OBCs. Here, $\Pi$ stands for the operator representing inversion symmetry, i.e., $\Pi=\sigma_{X_{4N+2}}$. This fundamental relationship establishes a noteworthy outcome, namely, the shared eigenvectors of both the Hamiltonian and $\Pi$, creating a well-defined inversion symmetry operator \cite{Inv_sym3,Bahari2019}
\begin{eqnarray}
   \langle \Tilde{\Psi}_{m}| \Pi |\Tilde{\Psi}_{m}\rangle=\pm1.
\end{eqnarray}
Therefore we can use the expectation value of this operator to define two invariants,
\begin{eqnarray}
    Z_1&=&\sum_{m=1}^{N_f} \langle\Tilde{\Psi}_{m}| \Pi |\Tilde{\Psi}_{m}\rangle,\\
    Z_2&=&\sum_{m=1}^{N_f} [(\langle \Tilde{\Psi}_{m}| \Pi |\Tilde{\Psi}_{m}\rangle-1)/2],
\end{eqnarray}
where $N_f$ is the number of states below Fermi energy $E_f=0$.
Here $Z_1$ counts the total parity of all states with ${\rm Re}[E]<0$, and $Z_2$ counts only the ones with negative parity.
Note that $Z_2$ may change with the system's size, and it is its jump that characterizes the change of a system's overall parity.

In the bottom two panels of Fig. \ref{fig6}, the values of $Z_1$ and $Z_2$ as a function of $\theta/\pi$ are depicted. In Fig. \ref{fig6}(a), both $Z_1$ and $Z_2$ jump discontinuously at the bulk gap closing points, simultaneously, revealing both the transitions of anti-$\mathcal{PT}$ symmetry and topological phases.
For the parameter regions where these two transitions separate, it is found that the topological phase transition is characterized by the jump of $Z_1$, and the anti-$\mathcal{PT}$ transition is characterized by the jump of $Z_2$, as displayed in Fig. \ref{fig6}(b). Finally, as shown in Fig. \ref{fig6}(c) with $\gamma=2$, $Z_1$ remains unchanged for all values of $\theta$, as now the system hosts only the anti-$\mathcal{PT}$ transition marked by edge-EPs and the jumps of $Z_2$.

\section {Edge modes in the time domain} \label{s5}

NH systems exhibit unique dynamics due to the existence of localized eigenstates with large imaginary energies. These states suggest novel propagation scenarios that dominate the system's behavior over long periods of time. Although the time evolution of such systems is non-unitary, any initial state can still be decomposed into a combination of different eigenstates. Therefore, we can design various schemes to test our model.

To investigate the behavior of the system under NH time evolution, we prepare two types of initial states, i.e., edge states and bulk states. By analyzing their time evolution, we aim to understand whether the initial states remain localized or spread throughout the system. We define the normalized spatial distribution of the final state,
\begin{eqnarray}\label{e1}
\rho_x(t)&=& |\Psi_f(x,t))|^2/\langle \Psi_f(t)|\Psi_f(t)\rangle,\\
|\Psi_f(t)\rangle&=&U(t)|\Psi_i\rangle,
\end{eqnarray}
where $\Psi_f(t)$ ($\Psi_i$) is the finial (initial) state and $U(t)=e^{-iHt}$ is the time evolution operator of the system after time $t$. 

In Figs. \ref{fig8} and \ref{fig9}, we simulate $\rho_x(t)$ as functions of position and time, for the commensurate and incommensurate case, respectively. These simulations were conducted with initial states prepared at distinct positions either within the bulk (left columns) or at the edges (right columns) of the system, and the parameters chosen for the system to host a pair of topological edge states with zero real-energy [as in Fig. \ref{fig4}(c) and \ref{fig4} (d)]. We find that for the commensurate case, the final state always evolves to the left edge of the system, regardless of where the initial state is placed, as shown in Fig. \ref{fig8}. In particular, as can be seen in Fig. \ref{fig8}(f), a state prepared at the right edge shows a clear diffusion even in a short period of time, despite the presence of a topological edge state localized at this edge. Physically, this is because the left edge state possesses a large positive imaginary energy and thus dominates the long-time dynamics, while the negative imaginary energy of the right edge state represents a strong dissipation, resulting in the dynamical instability of a state localized at this edge.

On the other hand, for the incommensurate case, topological edge states at both sides have the same positive imaginary energy, therefore they shall both have a significant impact on the dynamics. Indeed, we find that an inversion-symmetric initial state, whether placed in the bulk or at the edges, results in a balanced distribution localized at the two edges for the final state, as shown in Figs. \ref{fig9}(a) and \ref{fig9}(d). For non-symmetric initial states, the final states display a strong localization at only one of the two edges, depending on the distribution tendency of the initial states. This is shown in Figs. \ref{fig9}(b) and \ref{fig9}(c) for the initial state within the bulk states and in Figs. \ref{fig9}(e) and \ref{fig9}(f) for the initial state at the edge states.

Our investigation stands out for exploring a new type of edge states protected by anti-$\mathcal{PT}$ symmetry. This differs from the topological monomodes described in Refs. \cite{monomode1,monomode2}, where NH chiral symmetry supports the edge mode without relying on inversion symmetry. To elucidate the characteristics of edge states in our study, we analyze their dynamics. Particularly in the commensurate case, these states share similarities with the monomodes discussed in \cite{monomode1}. Furthermore, in the incommensurate case, we identify inversion symmetry as a critical factor in preserving spatially symmetric edge localization in the system, presenting a unique aspect of our work. Interestingly, the dynamics in the incommensurate case reveals a spontaneous breaking of inversion symmetry, contrasting with existing literature.

\section {Conclusions} \label{s6}

In conclusion, our investigation into the one-dimensional Su-Schrieffer-Heeger lattice with modulating non-Hermitian on-site potentials reveals intricate topological characteristics in both commensurate and incommensurate configurations. The study highlights the emergence of exceptional points (EPs) and their divergence from conventional topological phase transitions marked by the Berry phase, particularly in presence of imaginary on-site potential. This leads to the occurrence of edge states in regimes with trivial Berry phase and necessitates the development of new topological invariants for their characterization. Additionally, our research underscores the importance of inversion asymmetry, rather than anti-$\mathcal{PT}$ symmetry, in influencing the imaginary part of edge state energies. The findings not only shed light on the unique topological behaviors within non-Hermitian systems but also propose a novel approach to understanding phase transitions and dynamics of carriers, enriching our comprehension of the underlying physics in these systems.

Most realistic and applicable systems are finite, lacking bulk translational invariance, unlike ideal systems, which exhibit bulk translational symmetry. The exploration of incommensurate case in our model provides insights into the interplay between inversion symmetry and translational symmetry, offering a model that can be adapted to analyze similar systems supporting the fundamental symmetry, inversion symmetry. So our model and and the defined invariants can be studied in real platforms such as cold atom systems, photonic setups, and mechanical configurations.

It should be noted that our model with anti-$\mathcal{PT}$ symmetry can be realized in photonic systems with non-Hermitican character \cite{ANtiExp} where a topological phase transition was observed experimentally by employing bulk dynamics \cite{ANtiExp1}. In such systems, the existence of robust edge states has been demonstrated \cite{ANtiExp2,ANtiExp3}. Moreover, selective control and enhancement of
the topologically induced state in the SSH chain in a one dimensional microwave setup have been reported \cite{ANtiExp4}.

\section*{acknowledgment}
M.J. acknowledges the research council of University of Zanjan for the financial support. L.E.F.F.T. acknowledges financial support by FONDECYT (Chile) through grant 1211038 and of The Abdus Salam International Center for Theoretical Physics and Simons Foundation.
L.L. acknowledges support from National Natural Science Foundation of China (Grant No. 12104519) and the Guangdong Project (Grant No. 2021QN02X073).

\end{document}